\documentclass[a4paper,10.5pt,twocolumn,twoside]{article}

\usepackage{graphicx}
\usepackage{color}
\usepackage[a4paper, left=1.4cm, right=1cm, top=1.7cm, bottom=1cm]{geometry}
\usepackage[super,comma,sort&compress,square]{natbib}
\usepackage{amsmath}
\usepackage{float}
\usepackage{latexsym}
\usepackage{amssymb}

\usepackage{fancyhdr}
\usepackage[utf8x]{inputenc}
\usepackage[T1]{fontenc}

\begin{document}
\pagestyle{fancy}
\def\headrulewidth{0.5pt}
\def\footrulewidth{0pt}
\lhead{Applied Surface Science 302 (2014), 129 -- 133} 
\chead{}
\rhead{DOI: 10.1016/j.apsusc.2013.09.157}

\lfoot{} 
\cfoot{}
\rfoot{}

\twocolumn[
  \begin{@twocolumnfalse}
  {\huge \bf Determination of grain shape of laser-irradiated FePdCu thin alloy films}

  \hspace{1.1cm}
  \parbox{.87\textwidth}{
    \vspace{4ex}
    \Large \textsf{Marcin Perzanowski, Michal Krupinski, Arkadiusz Zarzycki, Yevhen Zabila, Marta~Marszalek}
    \vspace{1ex} \\
    \normalsize The Henryk Niewodniczanski Institute of Nuclear Physics, Polish Academy of Sciences, Radzikowskiego 152, 31-342 Krakow, Poland
    \vspace{1ex} \\
    \normalsize \text{email: Marta.Marszalek@ifj.edu.pl, Marcin.Perzanowski@ifj.edu.pl}

    \vspace{2ex} 
    \noindent
     \textbf{Abstract}: The irradiation with the 10~ns pulsed infrared Nd:YAG laser was applied to transform FePdCu multilayers into chemically ordered L1$_{0}$ phase. 
     The X-ray diffraction methods ($\Theta/2\Theta$ scan, $\psi$-scan, $\omega$-scan) were used to trace the presence of L1$_{0}$ phase after laser annealing with different number of pulses. 
     The size and shape of crystallites was determined depending on their orientation with respect to film plane. 
     The (111) oriented crystallites of constituent metals were built as coherent domains spreading through multilayers during deposition of films. 
     Laser annealing induced the transformation of multilayers to alloy, and the ordering of (111) oriented crystallites. 
     Simultaneously, the (002) oriented crystallites appeared confirming the transformation to L1$_{0}$ alloy.
     
     \vspace{2ex}
     DOI: 10.1016/j.apsusc.2013.09.157 
     
     \vspace{2ex}
     
    \vspace{3ex}
  }
  \end{@twocolumnfalse}
]

\section{Introduction}

\vspace{-0.4cm}
\indent
The microstructure evolution directly influences the physical, mechanical, electrical and also chemical properties of materials. 
In thin films, size effects related to structure may also induce more complex effects demanding special attention during experiment
and interpretation of the results. 
In particular, the thermal decomposition of metallic thin films and multilayers is a~field which allows to generate original microstructures such as non-equilibrium precipitates in composites\cite{1} or dispersed particles with different core and surface composition.\cite{2} 
The different means of annealing are used in research and industry, the most typical devices are ovens which allow the heating in atmosphere/air or flash lamp heating like Rapid Thermal Annealing. 
However, the other techniques are also developed like plasma arc heating, microwave and laser heating. 
Laser processing of materials became recently an important field of research due to its broad applications also in micromachining, patterning, thin film deposition and welding.\cite{3,4,5,6}

Here, we present the application of pulsed laser heating in phase transformation of FePdCu thin alloy films. 
The FePd alloy is promising candidate for Bit Patterned Media technology due to its strong perpendicular magnetic anisotropy.\cite{7} 
However, this feature appears only in alloys with L1$_{0}$ structure, epitaxial or polycrystalline with strong (001) texture.\cite{8} 
The tetragonal L1$_{0}$ structure with the alternating planes of Fe and Pd atoms is the anisotropic crystal with strong magnetocrystalline anisotropy. 
The direction of magnetization normal to the film surface and perpendicular magnetic anisotropy appear only if the [001] crystallographic direction is perpendicular to the sample plane. 
The thin alloy films can be fabricated by different deposition techniques like thermal evaporation, sputtering and molecular beam epitaxy either from composite alloys or by co-deposition, but as-grown film alloy deposited at room temperature on amorphous substrates crystallizes in disordered fcc structure. 
The formation of the L1$_{0}$ crystal structure occurs only after annealing at temperatures above 500$^{\circ}$C. 
Usually such annealing temperature is sufficient\cite{9,10} to fabricate the chemically ordered L1$_{0}$ alloy, but is not large enough to produce the alloy with a (001) texture, which is needed to obtain perpendicular magnetic anisotropy. 
To reduce the temperature of L1$_{0}$ structure formation and to induce (001) texture the FePd alloy was doped with Cu, as it was observed in our previous studies.\cite{11,12} 
We present here the studies of the laser annealing effect on the structure, surface morphology, grain size and shape of FePdCu thin
alloy films. 

\vspace{-0.5cm}
\section{Experimental details}
\vspace{-0.3cm}
\subsection{Samples preparation}

The [Cu(0.2 nm)/Fe(0.9 nm)/Pd(1.1 nm)]$_{10}$ multilayers were thermally evaporated on Si(100) substrates covered by a~native SiO$_{2}$ oxide layer. 
The thickness of Fe and Pd layers was chosen as corresponding to 48:52 atomic stoichiometry, necessary to obtain
the L1$_{0}$ structure. 
Before the deposition substrates were ultrasonically cleaned in acetone and ethanol and then rinsed in de-ionized
water. 
The evaporation was carried out in ultrahigh vacuum with the working pressure of the order of 10$^{-9}$~mbar. 
The layers were evaporated sequentially with rates of 0.5~nm/min for Pd and Fe, and 0.2 nm/min for Cu. 
The thickness of the layers was monitored in situ using quartz crystal microbalance. 
The chemical composition of films was confirmed by Rutherford Backscattering Spectrometry after deposition. 
The layer structure of the samples was investigated using X-ray reflectivity (XRR). 
In order to obtain information about thickness and roughness of the layers the data were fitted
using X’Pert Pro Reflectivity software, which takes the advantage of the Parrat algorithm.\cite{13} 
XRR data confirmed the nominal thicknesses of the films with accuracy of about 5\% with respect to the nominal thickness. 
XRR results also showed, that the average roughness of the single layer had a value of about 0.6~nm, which suggests a significant intermixing between the layers during the deposition process.

\vspace{-0.4cm}
\subsection{Pulsed laser annealing}

\vspace{-0.2cm}
In order to obtain FePdCu alloy laser annealing of the multilayers was carried out using multimode Quantel YG980 Nd:YAG
pulsed laser, first harmonic with wavelength of 1064 nm, the pulse duration time of 10 ns with repetition frequency of 10 Hz. 
The spatial divergence of the beam, which was not focused, was 0.07 rad and beam spot diameter measured at half of the maximal intensity was equal to 4.5~mm. 
The energy of the beam was set with Q-switch device with minimal delay step of 1~$\mu$s. 
Energy density of the beam was controlled using Coherent FieldMax II energy meter. 
The energy distribution in beam spot was not perfectly homogeneous, however, its random character for the following pulses should result in average uniform areal distribution of the deposited energy. 
In order to prevent the oxidation annealing process was conducted in flowing nitrogen atmosphere. 
Samples were irradiated with 1, 10, 100, and 1000 of laser pulses. 
Energy density of the single pulse was 235~mJ/cm$^{2}$, which corresponds to the surface peak temperature of approximately 500$^{\circ}$C. 
Maximum temperature was estimated by calculations based on heat transfer equation and the theory of pulsed laser annealing presented
in [14,15]. 
Absorption of the radiation by multilayer at 1064 nm, essential for temperature calculations, determined by optical measurements was equal to 28\%. 
According to this model the surface temperature after single pulse drops down to room temperature during the time between pulses. 
It is worth mentioning, that the model does not take into account the temperature dependencies of the material properties such as absorption, density, specific heat and thermal conductivity, and assumes, that between laser pulses they do not change. 
The peak temperature of 500$^{circ}$C was chosen basing on results of our previous investigations,\cite{9,11,12} which showed that it is the minimum temperature required to induce a~diffusion process, and to start transformation from multilayer system into ordered alloy.

\vspace{-0.4cm}
\subsection{X-ray diffraction experiments}

\vspace{-0.2cm}
The diffraction experiments were performed with PANalytical X’Pert Pro diffractometer using Cu $K_{\alpha}$ line ($\lambda=0.154$~nm), operating at 40 kV and 30 mA. 
Parallel incident beam was formed with a~parabolic graded W/Si mirror with an equatorial divergence less than 0.05$^{\circ}$. 
The incident beam optics consisted of 1/2$^{\circ}$ divergence slit and 0.04~rad Soller slits. 
In order to restrict the axial divergence of the beam the 4~mm wide mask was used. 
The signal was recorded with solid state stripe detector with a~flat graphite monochromator. 
Diffracted beam path was equipped with antiscatter slit with height of 8.7~mm and 0.04~rad Soller slits. 
The diffraction experiments were performed in the following geometries: $\Theta$/2$\Theta$ X-ray diffraction with scattering vector perpendicular to the surface plane (XRD), $\Theta$/2$\Theta$ X-ray diffraction with scattering vector tilted at angle with respect to the diffraction plane (XRD-$\psi$), and in $\omega$ geometry (rocking curve measurements) where angle
2$\Theta$ was kept constant, while the angle of incident beam varied. 
The XRD and XRD-$\psi$ measurements were performed in 2$\Theta$ range from 35$^{\circ}$ to 55$^{\circ}$ with instrumental step of 0.05$^{\circ}$. 
Time of the single measurement was approximately 24~h. 
Rocking curves were collected in the range of incident angles $\pm17^{\circ}$ around the angle 2$\Theta$. 
Instrumental step of rocking curves was equal to 0.5$^{\circ}$, total time of single measurement was approximately 12~h. 
Diffraction patterns were corrected for instrumental background. 
The contribution of the Cu $K_{\alpha 2}$ line was subtracted using Rachinger algorithm.\cite{16}

\vspace{-0.4cm}
\subsection{Morphology investigations}

\vspace{-0.2cm}
Scanning electron microscopy (SEM) experiments were done using Nova NanoSEM (FEI) equipment. 
The primary electron beam was accelerated with voltage of~10 keV. 
The spot size of the primary beam was approximately 3~nm. 
All of the micrographs were collected using secondary electrons. 
The images were taken in immersion mode with the through-the-lens (TLD) detector in order to improve the resolution. 
Experiments were carried out at pressure of 10$^{-6}$~mbar.

\vspace{-0.4cm}
\section{Results and discussion}

\vspace{-0.2cm}
\subsection{Surface morphology after laser \mbox{annealing}}

\begin{figure}[!bh]
\centering
\includegraphics[width=0.25\textwidth]{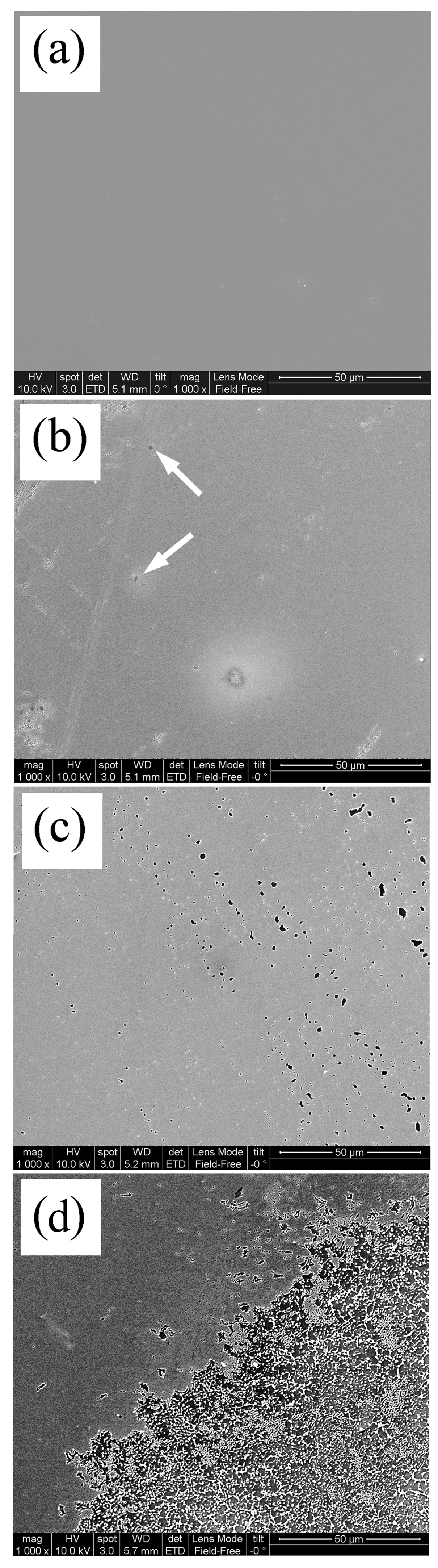}
\caption{SEM micrographs of as-deposited sample, similar to the system irradiated with single laser pulse (a), and irradiated with 10 (b), 100 (c), and 1000 (d) laser pulses. In (b) voids were indicated with white arrows.}
\label{fig1}
\end{figure}

\vspace{-0.1cm}
The topography of the samples after laser irradiation was investigated with SEM. 
The surface morphology of multilayers after single laser pulse (Fig.~\ref{fig1}a) did not exhibit any distinct features, similarly as the morphology of as-deposited sample. 
The micrographs showing surface morphology of samples annealed with 10, 100, and 1000 laser pulses are shown in Figs.~\ref{fig1}b -- d, respectively.
In the case of sample irradiated with 10 laser pulses annealing process led to creation of small number of voids in the surface with
the average area of single void about 0.22~$\mu$m$^{2}$. 
The diameter of a~single void was in the range from 0.4~$\mu$m to 1~$\mu$m in this case. 
After 100 laser pulses larger number of voids was observed and their average area increased to approximately 0.84~$\mu$m$^{2}$ and their diameter was in the range of 0.4 -- 2~$\mu$m. 
The irradiation with 1000 laser pulses caused partial degradation of the surface and exposure of the substrate. 
However, parts of the surface remained unaltered, which may be caused by inhomogeneous laser-beam profile.

\vspace{-0.5cm}
\subsection{Crystallographic structure}

\begin{figure}[!ht]
\centering
\includegraphics[width=0.5\textwidth]{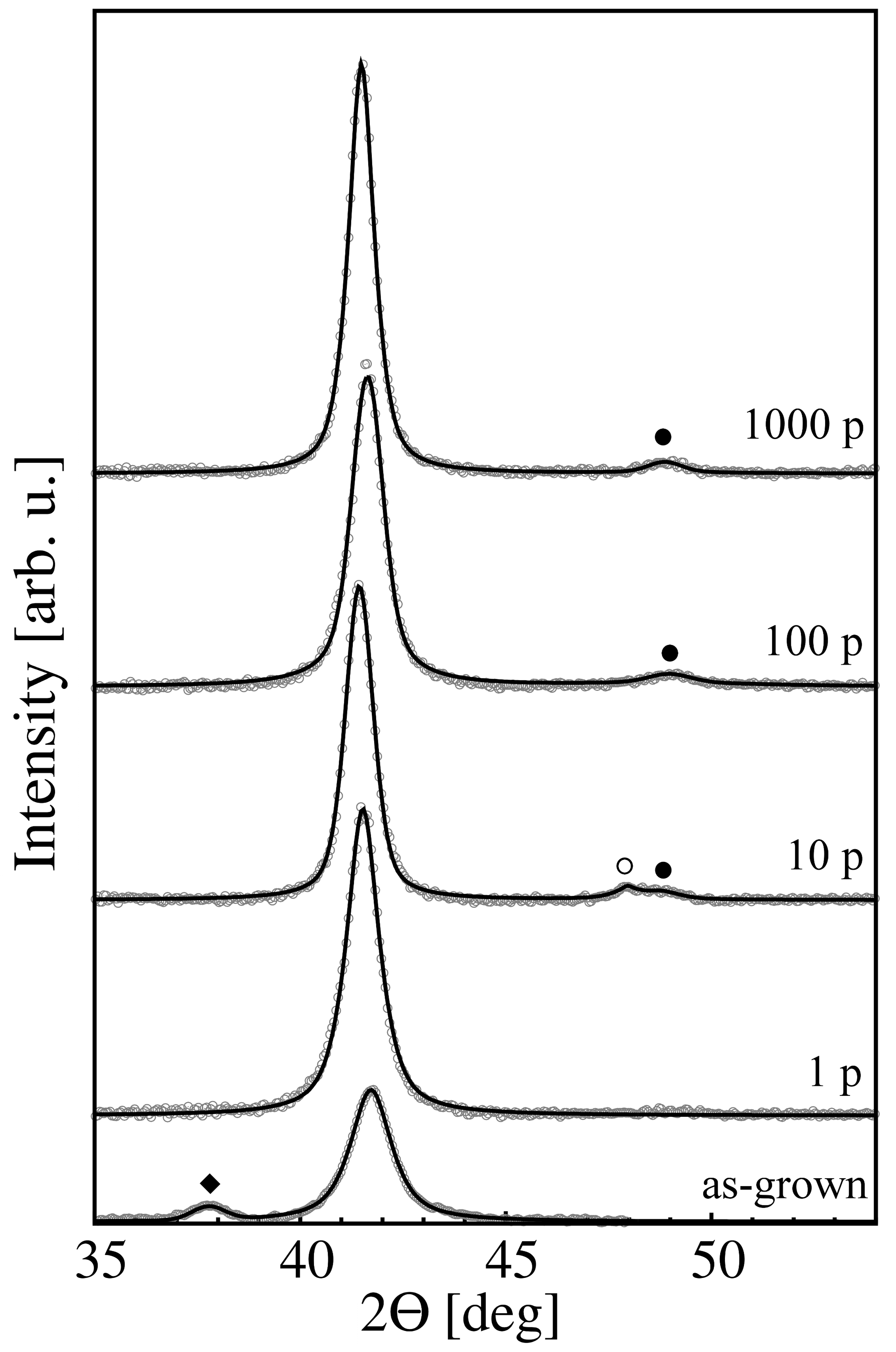}
\caption{X-ray diffraction patterns of as-grown multilayer and after irradiation with 1, 10, 100, and 1000 laser pulses. For as-grown sample the −1$^{\mathrm{st}}$ order satellite was marked by square. 
The L1$_{0}$ (002) reflections were indicated with full circles, L1$_{0}$-(200) reflection was marked by open circle. 
Solid lines are fit of the pseudo-Voigt function. 
Patterns were vertically shifted for clarity.}
\label{fig2}
\end{figure}

\vspace{-0.1cm}
The crystallographic structure of the as-grown multilayer and irradiated samples was investigated by XRD measurements, the
results are shown in Fig.~\ref{fig2}. 
In the pattern of as-grown sample two reflections at $2\Theta = 37.6^{\circ}$ and $41.7^{\circ}$ are observed. 
Both of them come from periodic multilayer superstructure and are satellites of -1$^{\mathrm{st}}$ and 0$^{\mathrm{th}}$ order.\cite{17} 
The lack of +1$^{\mathrm{st}}$ order satellite peak as well as higher order satellites is connected with intermixing between layers inducing perturbation of multilayer structure, proved by XRR. 
The Cu/Fe/Pd trilayer period, calculated from the distance between center and -1$^{\mathrm{st}}$ satellite peak, is equal to $2.41 \pm 0.14$~nm and stays in agreement with XRR data ($2.24 \pm 0.12$~nm) and nominal multilayer period (2.2~nm). 
The pattern collected for samples irradiated with single pulse shows only one reflection at $2\Theta = 41.6^{\circ}$. 
The disappearance of -1$^{\mathrm{st}}$ order satellite, in comparison to the as-grown system, is the evidence of decomposition of the multilayer structure and onset of the alloying process. 
It should be emphasized, that the change in crystal structure occurred after only one laser pulse. 
Taking into account the presence of only one diffraction peak and the fact, that FePd L1$_{0}$ and fcc structures have their (111) reflections nearly at the same angular position about 41.6$^{\circ}$,\cite{18} the crystallographic structure cannot be determined in this case.

In patterns obtained for samples irradiated with 10, 100, and 1000 laser pulses, two Bragg reflections at $2\Theta = 41.6^{\circ}$ and $48.9^{\circ}$ were recorded. 
These peaks were identified as (111) and (002) reflections of L1$_{0}$-ordered FePdCu alloy. 
For sample irradiated with 10 laser pulses the additional reflection at 47.5$^{\circ}$, identified as (200) peak of L1$_{0}$ structure, is observed. 
From Bragg equation and angular peak positions the lattice parameters $a$ and $c$ of the L1$_{0}$-ordered FePdCu alloy were determined, with the values of $a = 0.380 \pm 0.001$~nm and $c = 0.371 \pm 0.001$~nm. 
The lattice distortion $c/a = 0.98 \pm 0.01$ is in agreement with the value for bulk alloy (0.97).\cite{18} 
However, it is smaller than reported for rapidly annealed FePdCu alloy.\cite{12} 
The same authors reported, that for well-ordered FePd alloys Cu admixture causes decrease of $c$ value with simultaneous increase of $a$ value, in comparison to the bulk values.

In order to determine the crystallographic texture of irradiated alloys the $\omega$-scans were performed around (111) and (002) reflections of L1$_{0}$ structure. 
\begin{figure}[!h]
\centering
\includegraphics[width=0.45\textwidth]{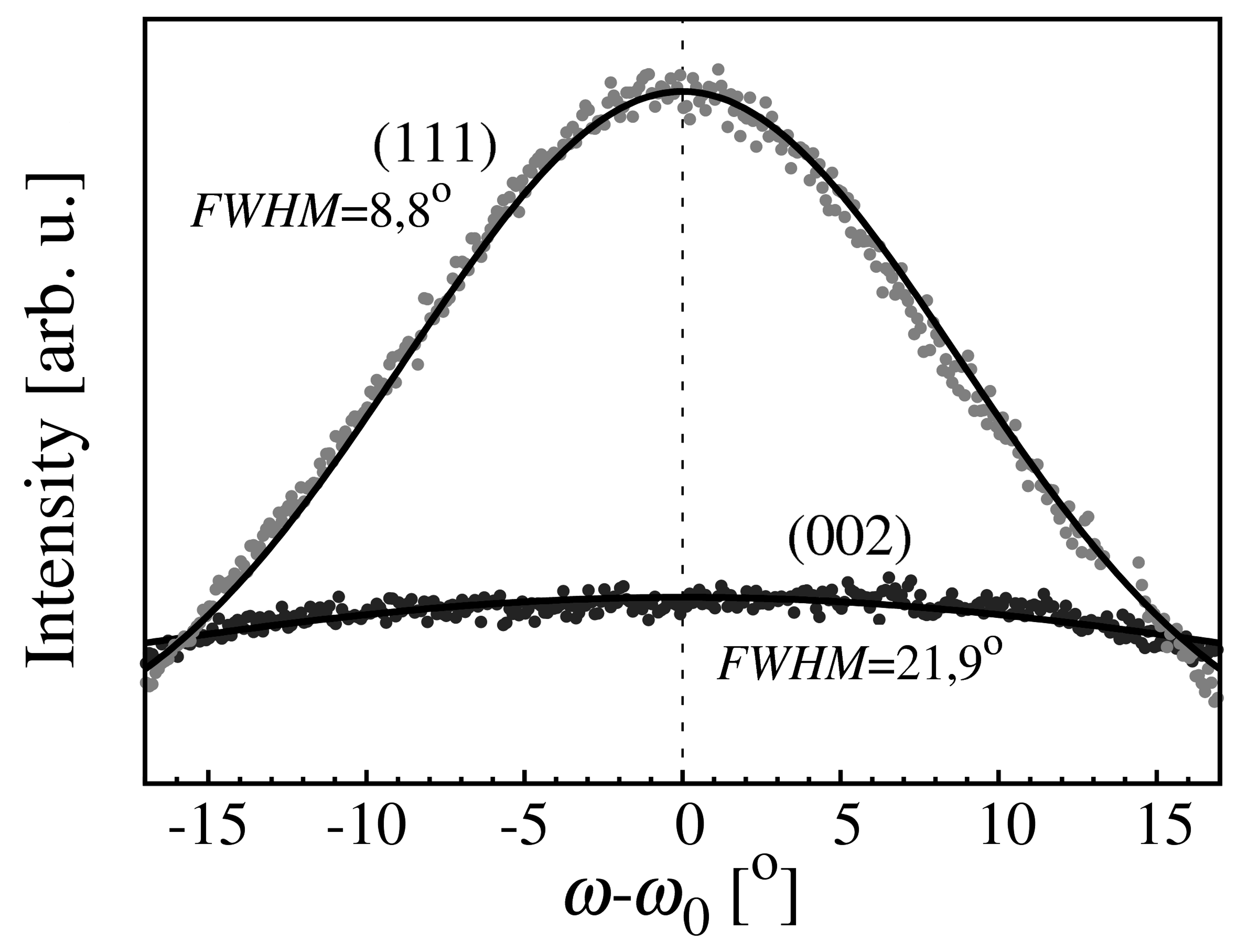}
\caption{Rocking curves measured for L1$_{0}$ (111) and (002) reflections of sample irradiated with 100 laser pulses.}
\label{fig3}
\end{figure}
Our previous studies showed, that FePdCu thin alloy films on Si(100) substrate revealed the fiber texture of [111] and [001] crystallites.\cite{12} 
It manifested in the cylindrical symmetry of grains oriented around the axis normal to film plane. 
Therefore, the result of $\omega$-scan did not depend on the polar angle at which the sample was placed with respect to the incident X-ray beam. 
The $\omega$-scans measured for sample irradiated with 100 laser pulses are presented in Fig.~\ref{fig3}. 
The $\omega$-scan for (002) reflection has FWHM width of about 22$^{\circ}$, which indicates a~nearly random distribution of spatial orientation of (002) crystallites. 
The intensity of rocking curve measured for (111) reflection has distribution with FWHM equal to 8.8$^{\circ}$. 
The lower value of FWHM for (111) rocking curve suggests a weak (111) crystallographic texture. 

\vspace{-0.4cm}
\subsection{Grain size and shape}

\begin{figure}[!h]
\centering
\includegraphics[width=0.5\textwidth]{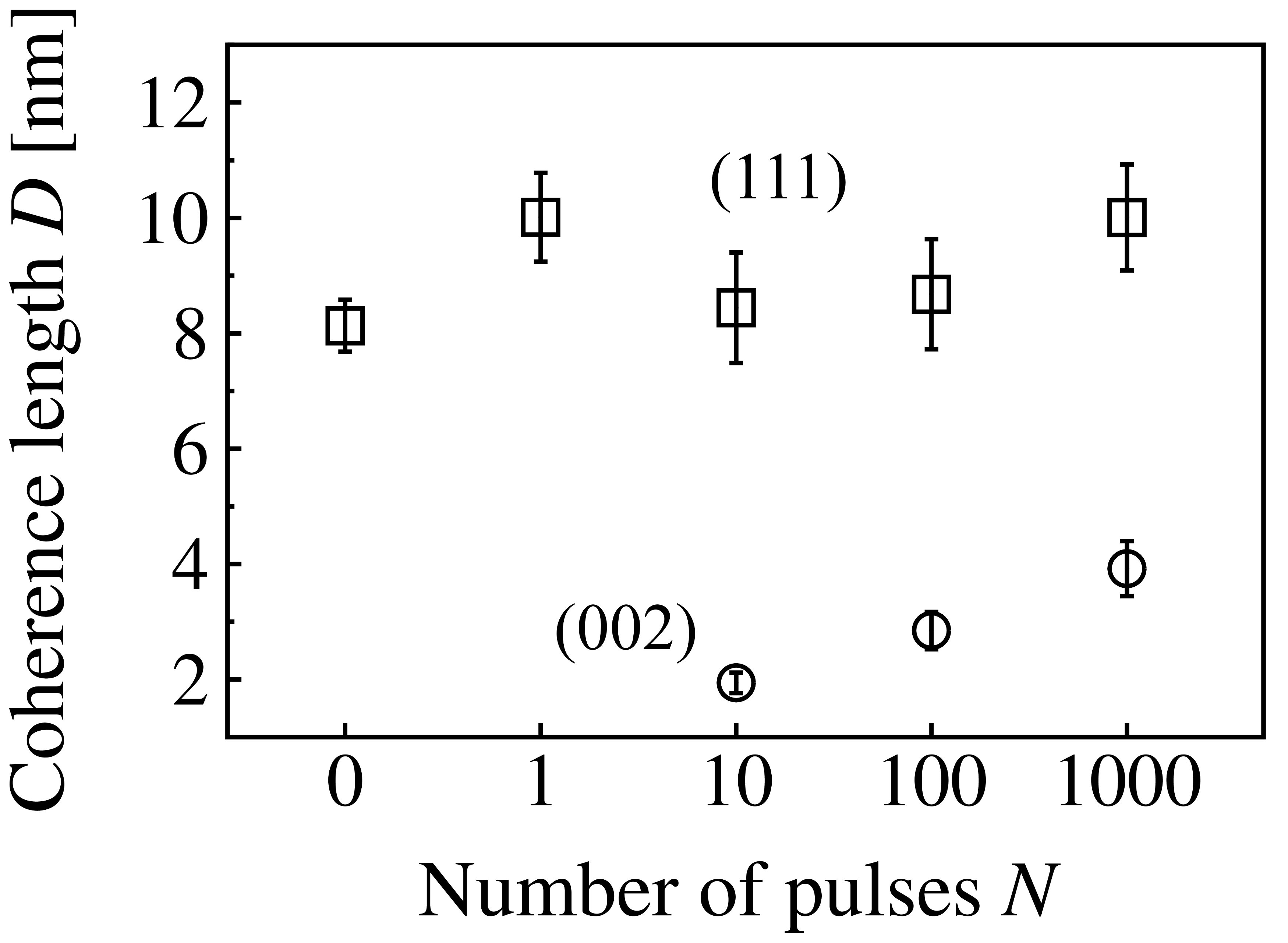}
\caption{Coherence length calculated from L1$_{0}$ (111) and (002) reflections as a function of number of laser pulses.}
\label{fig4}
\end{figure}

\vspace{-0.2cm}
Grain size investigations based on the XRD measurements, were performed using single line profile analysis method.\cite{16,19} 
The (111) and (002) reflections of L1$_{0}$ structure were fitted by pseudo-Voigt function in order to obtain the values of FWHM of the Cauchy term. 
Using Scherrer equation and the values of FWHM the vertical coherence lengths $D$ were determined. 
The value of dimensionless Scherrer factor was equal to 0.95, the most often applied in studies of thin alloy films. 
Values of the FWHM were corrected for instrumental peak broadening of 0.05$^{\circ}$. 
The results of the calculations as a~function of the number of laser pulses, are shown in Fig.~\ref{fig4}.

The appearance of the peak at approximately 41.6$^{\circ}$ in each XRD pattern clearly indicates, that in each sample the coherent domains with average interplanar distance of 0.217~nm are present. 
The value of their coherence length does not change with increasing number of laser pulses and equals to approximately 9~nm, which
is nearly half of the total thickness of the film. 
The (111)-oriented grains arise from coherent domains of the initial multilayer structure, with the transitional stage of undetermined crystallographic structure seen after single laser pulse irradiation. 
Lack of change of grain vertical size may suggests, that the pulsed laser annealing induces the ordering process only within these particular coherent domains. 
It is also possible, that laser irradiation leads to the grain growth in lateral direction, which cannot be seen by specular XRD measurement, since this method does not allow to study structural properties along direction different than normal to the surface plane.

The L1$_{0}$ (002)-oriented crystallites are formed as a~result of laser annealing. 
The irradiation rapidly increases the temperature, which stimulates diffusion and crystallization processes. 
The smallest (002)-oriented grains, having vertical size of 2~nm, were observed after 10 laser pulses. 
With increasing number of applied pulses their vertical size started to extend up to 4~nm after irradiation with 1000~shots.

\begin{figure}[!h]
\centering
\includegraphics[width=0.5\textwidth]{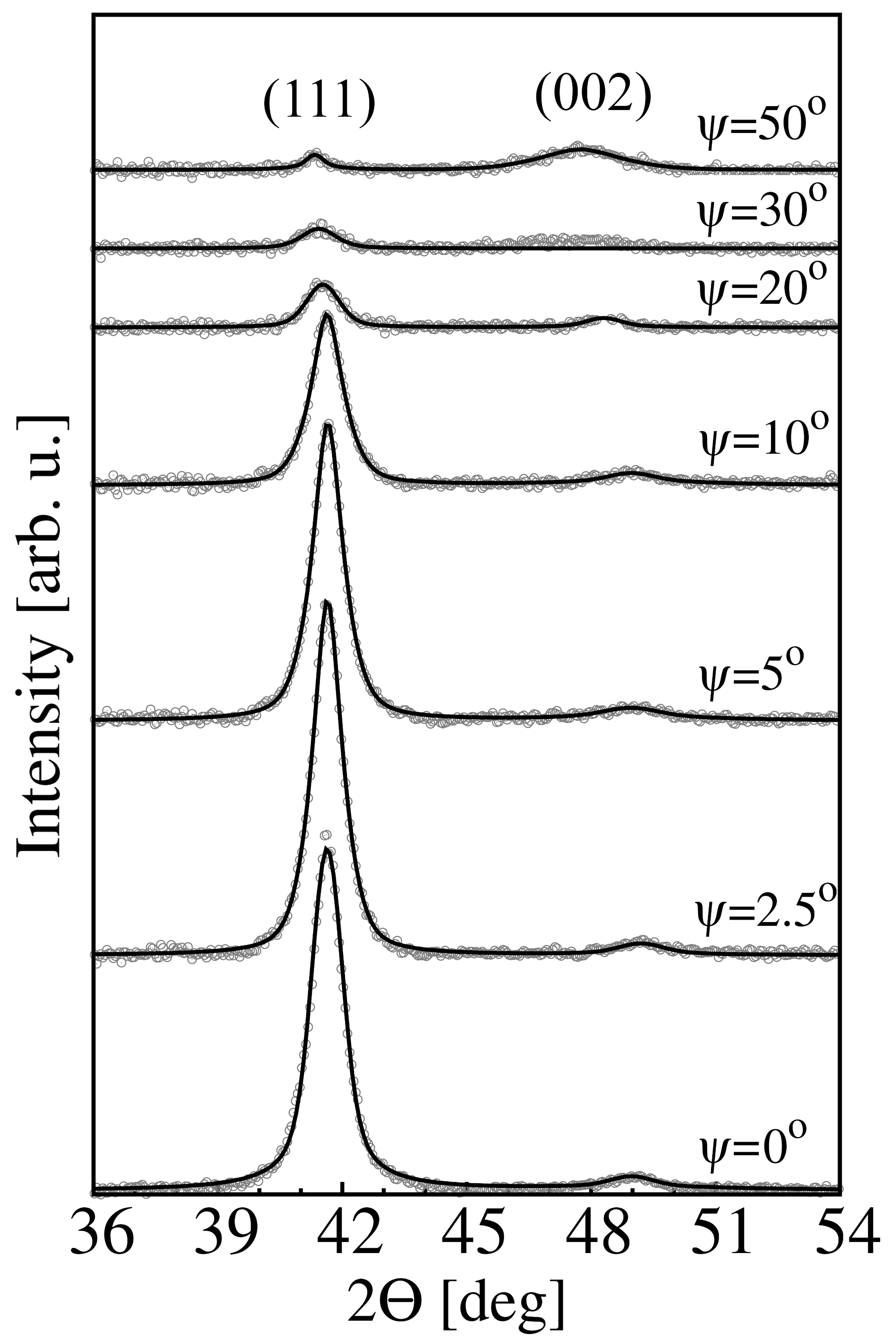}
\caption{X-ray diffraction patterns of sample irradiated with 100 laser pulses measured for different tilt angle $\psi$. 
Solid lines are fits of the pseudo-Voigt function. 
Patterns were vertically shifted for clarity.}
\label{fig5}
\end{figure}

The shape of crystallites was determined using XRD-$\psi$ measurements. 
The patterns obtained for sample irradiated with 100 laser pulses are shown in Fig.~\ref{fig5}. 
In the XRD-$\psi$ method the scattering vector is directed at angle different than normal to the surface plane, which allows to obtain information about coherence length in different spatial directions.\cite{20} 
In XRD-$\psi$ patterns (111) and (002) L1$_{0}$ reflections were observed for each tilt angle $\psi$. 
The peak positions are shifted with increasing value of tilt angle toward lower angles 2$\Theta$, which is connected to the increase of the interplanar distance between crystallographic planes and indicates nonuniform spatial distribution of stress in the sample.\cite{16} 
Changes in amplitudes of (111) and (002) reflections are in agreement with rocking curve scans presented in previous paragraph, and confirm the lack of sharp crystallographic texture.

\begin{figure}[!h]
\centering
\includegraphics[width=0.5\textwidth]{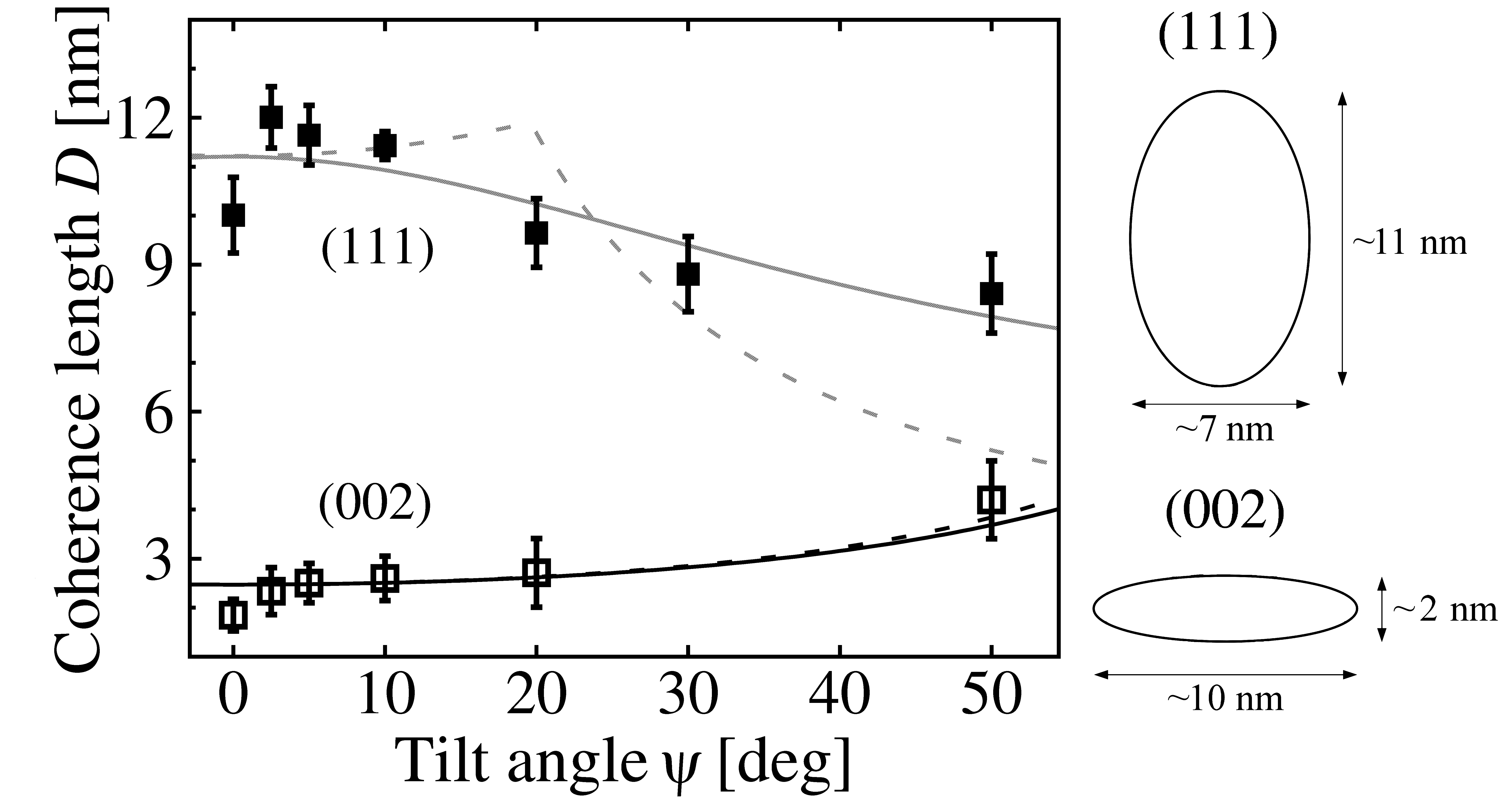}
\caption{Coherence length calculated from L1$_{0}$ (111) and (0 0 2) reflections (open squares and full squares, respectively) for sample irradiated with 100 laser pulses as the function of tilt angle~$\psi$. 
Dashed and solid lines are fits of functions corresponding to rectangular and elliptical shapes of crystallographic grains, respectively.}
\label{fig6}
\end{figure}

Applying the same calculation procedure as described above, the values of coherence length for (111) and (002) grains were
obtained and they are plotted against the values of tilt angle (Fig.~\ref{fig6}).
It was found, that coherence length for (111) grains decreases with increasing tilt angle, while the rise of $D$ values with increasing angle was observed for (002) crystallites. 
In order to determine both vertical and lateral sizes, and the shape of crystallites the obtained $D(\psi)$ dependencies were fitted with functions corresponding to rectangular and elliptical shape. 
The schematic drawings of (111) and (002) crystallites sizes and shapes, determined from the fits, are presented in Fig.~\ref{fig6}. 
It is seen, that (111) grains have shape elongated in the direction perpendicular to the surface plane, while (002) crystallites are elongated along direction parallel to the surface. 
In case of (111) crystallites the elliptical function gave better fit than rectangular. 
For (002) grains both functions gave the same fit quality. 
For estimation of lateral size of crystallites the values obtained from elliptical function fit were taken, and the lateral size
was interpolated for $\psi=90^{\circ}$.

Vertical elongation of (111) crystallites agrees with the model developed for FePdCu alloys annealed using Rapid Thermal Annealing method.\cite{11} 
The (111) oriented grains grow already during the deposition of films forming a~coherent domains through the multilayer. 
The annealing induces recrystallization process in these coherent domains leading to partial ordering and L1$_{0}$ phase creation. 
Assuming fiber orientation of crystallites (see Section~3.2), (11`)-oriented grains have circular lateral cross-section and therefore can be visualized in three-dimensional space as ellipsoids with average volume of about 270~nm$^{3}$. 
Laser irradiation initiates difussion process and in result (002)-oriented grains appear in spaces between (111) coherent domains. 
We believe that (002) grains can have oblate shape of disks with height of $2.5 \pm 0.2$~nm and diameter of approximately $10 \pm 2$~nm and average volume of about $195 \pm 15$~nm$^{3}$. 
The increase of their vertical size with increasing number of laser pulses (see Fig.~\ref{fig4}) indicates their continuous growth resulting from annealing, whereas the (111)-oriented grains are mostly transformed into L1$_{0}$ structure.

\vspace{-0.5cm}
\section{Conclusions}

\vspace{-0.2cm}
In this paper we presented studies of crystallographic structure, grain sizes and grain shapes of laser-irradiated FePdCu thin alloy
films. 
Pulsed laser annealing led to transformation from multilayer structure into L1$_{0}$-ordered alloy with weak (111) crystallographic
texture. 
It was found that increase of the number of laser pulses increases amount of cracks on the surface, and causes partial degradation of films for 1000 pulses. 
The XRD and XRD-$\psi$ measurements showed, that L1$_{0}$ (111)-oriented grains are formed from coherent domains present in multilayer structure and their vertical size does not depend on number of laser pulses. 
The (111) crystallites have elliptical shape elongated in direction perpendicular to surface. 
Laser annealing induces ordering process leading to formation of partially ordered L1$_{0}$ phase. 
The crystallization of (002)-oriented grains starts after irradiation with 10 laser pulses, and their vertical size increases with increasing number of laser pulses. 
This crystallites have oblate shape of disks with the longer axis of symmetry parallel to the sample surface. 
The effect of the residual stress resulting from the ultrafast laser annealing of multilayers, not included into this discussion, will be analyzed in the forthcoming paper.

\noindent
\vspace{0.3cm}

\textbf{Acknowledgements:} This work was supported by Polish National Science Center with Contracts Nos. 2012/07/N/ST8/00533 and 2012/05/B/ST8/01818. The authors are also grateful to Fabian Ganss from Chemnitz University of Technology for the help in SEM measurements.

\bibliographystyle{plainnat}

\begin{thebibliography}{10}

\bibitem{1} W.~Y.~Zhang, H.~Shima, F.~Takano, H.~Akinaga, S.~Nimori, Determining the low-coercivity temperature coefficient in FePt (fct)/FePt (fcc) nanocomposite films, J.~Phys.~Conf.~Ser. 200 (2010), 072106.

\bibitem{2} J.~H.~Hodak, A.~Henglein, M.~Giersig, Laser-induced inter-diffusion in AuAg core–shell nanoparticles, J. Phys. Chem. B 104 (2000), 11708 -- 11718.

\bibitem{3} U.~Klotzbach, A.~F.~Lasagni, M.~Panzner, V.~Franke, Laser Micromachining Fabrication and Characterization in the Micro-nano Range, Advanced Structured Materials, 10, Springer-Verlag, Berlin Heidelberg, 2011.

\bibitem{4} Y.~Zabila, M.~Perzanowski, A.~Dobrowolska, M.~Kac, A.~Polit, M.~Marszalek, Direct laser interference patterning: theory and application, Acta Phys. Pol. A 115 (2009), 591 -- 593.

\bibitem{5} M.~Ozegowski, K.~Meteva, S.~Metev, G.~Sepold, Pulsed laser deposition of multicomponent metal and oxide films, Appl. Surf. Sci. 138 -- 139 (1999), 68 -- 74.

\bibitem{6} F.~Bachmann, Industrial applications of high power diode lasers in materials processing, Appl. Surf. Sci. 208 -- 209 (2003), 125 -- 136.

\bibitem{7} T.~Schied, A.~Lotnyk, C.~Zamponi, L.~Kienle, J.~Buschbeck, M.~Weisheit, B.~Holzapfel, L.~Schultz, S.~F\"ahler, Fe-Pd thin films as a model system for self-organized exchange coupled nanomagnets, J. Appl. Phys. 108 (2010), 033902.

\bibitem{8} J.~R.~Skuza, C.~Clavero, K.~Yang, B.~Wincheski, R.~A.~Lukaszew, Microstructural, magnetic anisotropy, and magnetic domain structure correlations in epitaxial FePd thin films with perpendicular magnetic anisotropy, IEEE Trans. Magn. 46
(2010), 1886.

\bibitem{9} M.~Perzanowski, Y.~Zabila, J.~Morgiel, A.~Polit, M.~Krupinski, A.~Dobrowolska, M.~Marszalek, AFM, XRD and HRTEM studies of annealed FePd thin films, Acta Phys. Pol. A 117 (2010), 423 -- 426.

\bibitem{10} D.~E.~Laughlin, K.~Srinivasan, M.~Tanase, L.~Wang, Crystallographic aspects of L1$_{0}$ magnetic materials, Scrip. Mater. 53 (2005), 383.

\bibitem{11} M.~Krupinski, M.~Perzanowski, A.~Polit, Y.~Zabila, A.~Zarzycki, A.~Dobrowolska, M.~Marszalek, X-ray absorption fine structure and X-ray diffraction studies of crystallographic grains in nanocrystalline FePd:Cu thin films, J. Appl. Phys. 10 (2011), 064306.

\bibitem{12} M.~Perzanowski, Y.~Zabila, M.~Krupinski, A.~Zarzycki, A.~Polit, M.~Marszalek, Chemical order and crystallographic texture of FePd:Cu thin alloy films, J. Appl. Phys. 11 (2012), 074301.

\bibitem{13} L.~G.~Parrat, Surface studies of solids by total reflection of X-rays, Phys. Rev. 9 (1954), 359.

\bibitem{14} Z.~L.~Liau, B.~Y.~Tsaur, J.~W.~Mayer, Laser annealing for solid-phase thin-film reactions, Appl. Phys. Lett. 3 (1979), 221. 

\bibitem{15} Y.~Inaba, I.~Zana, C.~Swartz, Y.~Kubota, T.~Klemmer, J.~W.~Harrell, G.~B.~Thompson, Time–temperature–transformation measurements of FePt thin films in the millisecond regime using pulse laser processing, J. Appl. Phys. 10 (2010), 103907.

\bibitem{16} M.~Birkholtz, Thin Film Analysis by X-ray Scattering, Wiley-VCH Verlag GmbH \& Co. KGaA, Weinheim, 2006.

\bibitem{17} N.~Zotov, J.~Feydt, T.~Walther, A.~Ludwig, Structure of PtFe/Fe double-period multilayers investigated by X-ray diffraction, reflectivity, diffuse scattering and TEM, Appl. Surf. Sci. 253 (2006), 128 -- 132.

\bibitem{18} W.~B.~Pearson, A~Handbook of Lattice Spacings and Structures of Metals and Alloys, Pergamon Press, London, 1967. 

\bibitem{19} E.~J.~Mittemeijer, U.~Welzel, The state of the art of the diffraction analysis of crystallite size and lattice strain, Z. Kristallogr. 223 (2008), 552 -- 560.

\bibitem{20} F.~Jim\'enez-Villacorta, A.~M\'unoz-Martin, C.~Prieto, X-ray diffraction and extended X-ray absorption fine-structure characterization of nonspherical crystallographic grains in iron thin films, J. Appl. Phys. 96 (2004), 6224 -- 6229.


\end{thebibliography}
\vspace{1.1ex}
\begin{center}
 $\star$ $\star$ $\star$
\end{center}

\vspace{-9ex}

\setlength{\bibsep}{0pt}
\renewcommand{\bibnumfmt}[1]{$^{#1}$}

\end{document}